\documentclass[journal]{IEEEtran}
\IEEEoverridecommandlockouts
\usepackage{algorithm}
\usepackage{algorithmic}
\usepackage{tabularx}
\usepackage{pifont} 
\usepackage{ragged2e}

\usepackage{booktabs}

\usepackage{array}
\usepackage{etoolbox}

\usepackage[pdftex]{graphicx}
\graphicspath{{../pdf/}{../jpeg/}}
\DeclareGraphicsExtensions{.pdf,.jpeg,.png}

\usepackage[cmex10]{amsmath}
\usepackage{array}
\usepackage{mdwmath}
\usepackage{mdwtab}
\usepackage{eqparbox}
\usepackage{stfloats} 

\usepackage{url}
\usepackage{hyperref} 
\usepackage{orcidlink} 
\usepackage{cite}
\usepackage{mathrsfs}
\usepackage{amsmath,amssymb,amsfonts}
\usepackage{textcomp}

\usepackage{lipsum}
\usepackage{caption}
\usepackage{subcaption}
\usepackage{multirow}
\usepackage{nomencl}
\makenomenclature
\begin{document}

\bstctlcite{IEEEexample:BSTcontrol}
    \title{A Physics Informed Learning-Augmented Framework for Grid-Aware P2P Energy Trading}
  \author{

 Devangi,\,~\IEEEmembership{Graduate Student Member,~IEEE},  
    Ankit Singhal,\,~\IEEEmembership{Senior Member,~IEEE},  
    Yashasvi Bansal,\,~\IEEEmembership{Senior Member,~IEEE}  \\
    Indian Institute of Technology Delhi, Hauz Khas, 110016\\
    Emails:- eez248399@ee.iitd.ac.in, sankit@ee.iitd.ac.in, yashasvi@ee.iitd.ac.in
}


\maketitle

\begin{abstract}
Distribution networks are transitioning from passive to active systems, driven by the large-scale proliferation of distributed energy resources (DERs) at the grid edge. Peer-to-Peer (P2P) energy trading has emerged as a transformative paradigm that enables energy transactions among prosumers, aggregated here as microgrids (MG). Ensuring that these transactions remain physically feasible within the distribution network is a fundamental prerequisite for any credible P2P framework. This is typically handled by distribution system operators (DSO) through bilevel coordination in distributed P2P architectures, wherein MGs optimize trades independently of network constraints, and the DSO subsequently corrects the infeasible outcomes. However, such sequential coordination prevents MGs from anticipating network feasibility during P2P market clearing, potentially leading to substantial post-market corrections, altered trading outcomes, and repeated MG–DSO interactions. To this end, this paper proposes a physics-informed learning-augmented P2P–DSO framework in which a transformer-based model is trained to predict the DSO's response to proposed P2P trades, with network feasibility embedded directly in the training objective. Embedding this model within P2P market clearing algorithm allows MGs to anticipate the DSO's response and refine their trading decisions during trading, before submission. The proposed framework is validated on the modified IEEE 69-bus distribution system with interconnected MGs. Relative to conventional bilevel DSO correction, case studies show that the proposed framework recovers 40.90\% of P2P market utilization and 22.00\% of economic transactions, while empirically eliminating network constraint violations over a full year of operation and substantially reducing computational and communication overhead.
\end{abstract}



%
\IEEEpeerreviewmaketitle

\renewcommand{\nomgroup}[1]{%
\item[\itshape
\ifstrequal{#1}{I}{Indices and Sets}{%
\ifstrequal{#1}{V}{Variables}{%
\ifstrequal{#1}{P}{Parameters}{%
\ifstrequal{#1}{F}{Functions}{%
}}}}%
]}


\nomenclature[I]{$t$}{Time index}
\nomenclature[I]{$i$}{Microgrid index}
\nomenclature[I]{$n,\ell$}{Node and line index of the distribution network }
\nomenclature[I]{$k$}{Transformer encoder layer index}

\nomenclature[P]{$\alpha_i,\beta_i$}{Marginal and diminishing coefficient of MG $i$}
\nomenclature[P]{$a_i,b_i$}{Quadratic and linear generation cost coefficient}
\nomenclature[P]{$\eta_c,\eta_d$}{Charging and discharging efficiency of battery}

\nomenclature[P]{$\mathcal{L}^{min}_{i,t}, \mathcal{L}^{max}_{i,t}$}{Minimum and maximum power load at time $t$}
\nomenclature[P]{$\mathcal{G}^{max}_{i,t}$}{Maximum power generated by DG at time $t$}
\nomenclature[P]{$\mathcal{R}_{i,t}$}{Power generated by renewable at time $t$}

\nomenclature[P]{$SOC_{i}^{\max}$}{Maximum state of charge}
\nomenclature[P]{$\lambda_{d}$}{Battery degradation cost coefficient}
\nomenclature[P]{$SOC_{i}^{\min}$}{Minimum state of charge}

\nomenclature[P]{$\mathcal{P}_{i,max}$}{Maximum battery power capacity}

\nomenclature[P]{$r_{\ell},x_{\ell}$}{Resistance and reactance of line $\ell$}

\nomenclature[P]{$c^{ls}$}{Cost coefficient for line losses }
\nomenclature[P]{$\lambda_{corr}$}{Penalty factor for correction}
\nomenclature[P]{$\bar{v}_n, \underline{v}_n$}{Maximum and minimum voltage at node $n$}
\nomenclature[P]{$\bar{c}_\ell$}{Maximum squared current magnitude in line $\ell$}
\nomenclature[P]{$\bar{P}^{gen}_n, \bar{Q}^{gen}_n$}{Maximum active and reactive power at node $n$}

\nomenclature[V]{$P_\ell,Q_\ell$}{Active/reactive power flow on line $\ell$}
\nomenclature[V]{$\mathcal{P}^{gen}_n,\mathcal{Q}^{gen}_n$}{Active/reactive power generated at node $n$}
\nomenclature[V]{$\mathcal{G}_{i,t},\mathcal{L}_{i,t}$}{Power generated by DG and load at time $t$}
\nomenclature[V]{$\mathcal{P}^{dch}_{i,t},\mathcal{P}^{ch}_{i,t}$}{Discharging/charging power (BESS) at time $t$}
\nomenclature[V]{$\pi_t$}{Internal P2P price at time $t$}
\nomenclature[V]{$SOC_{i,t}$}{State of charge at time $t$}
\nomenclature[V]{$c_\ell,v_n$}{Squared current of line $\ell$ and voltage at node $n$}
\nomenclature[V]{$\mathcal{P}^{load}_n,\mathcal{Q}^{load}_n$}{Active and reactive power load at node $n$}
\nomenclature[F]{$U(\cdot)$}{Utility function}
\nomenclature[F]{$C(\cdot)$}{Cost of generation function}
\nomenclature[F]{$W(\cdot)$}{Payoff function}
\nomenclature[F]{$\mathcal{L}(\cdot)$}{Mean Squared Error (MSE) loss function}
\nomenclature[F]{$\mathcal{F}(\cdot)$}{DSO objective function}


\renewcommand{\nomlabel}[1]{\makebox[1.8cm][l]{#1}}
\input{output.nls}

\section{Introduction}
Modern distribution systems increasingly host distributed energy resources (DERs), transforming the grid edge into an active marketplace for energy and flexibility \cite{muhtadi2021distributed}. Peer-to-Peer (P2P) energy trading channels this capability by letting prosumers exchange energy bilaterally, enhancing renewable utilization while reducing costs and grid stress \cite{tushar2020peer}. Consequently, P2P energy trading among interconnected MicroGrids (MGs) has emerged as a key paradigm for enabling local balancing and flexible energy exchange \cite{wang2021stochastic}. Game-theoretic frameworks are widely used to model P2P among MGs and enable decentralized coordination \cite{belgioioso2022operationally}. Depending on the coordination, they are classified as cooperative or non-cooperative \cite{amin2020converging}.

However, these trades, executed at the market layer, directly affect the physical network, which can violate voltage limits, line loading limits, and generator output ratings \cite{paudel2020peer}. Such violations risk equipment damage, protection trips, line overloads, and even large-scale outages \cite{8782819}. 
Several studies incorporate physical network constraints into P2P market formulations. Network-aware P2P trading has been addressed through coordinated energy management \cite{zhang2020coordinated}, network-aware fairness \cite{zamani2024network}, and network usage charges in peer matching \cite{trivedi2025p2p}, while \cite{feng2022peer} employs a co-simulation platform integrating distribution network models with P2P trading mechanisms. However, these papers largely overlook the economic  role of the DSO, leading to an incomplete representation of market-network interactions.
Beyond ensuring physical network constraints, the DSO also generates wheeling-fee revenue from P2P exchanges, supporting network operation and expansion. Therefore, enabling more feasible P2P trades supports both system reliability and the DSO’s economic objectives \cite{sheng2022incorporating}.
In several formulations, the DSO is modeled as a centralized authority with full control over MG scheduling and trade outcomes \cite{zhong2020cooperative}, \cite{angaphiwatchawal2022centralized}. Such centralized structures compromise MG autonomy, require extensive access to internal MG information. On the other hand, distributed coordination schemes such as those in \cite{zou2022peer}, \cite{yan2020distribution} rely on frequent exchange of detailed operational data, including marginal costs, power schedules, dual variables, among MGs and the DSO. Such extensive communication increases cyber risks and limits practical scalability. Moreover, the iterative coordination between MGs and DSOs suffers from additional limitations. Specifically, network-enforced corrections can significantly alter market outcomes, while repeated interactions can increase the coordination overhead and computational burden, ultimately eroding the economic benefits of P2P trading. 

\begin{figure*}[b]
\vspace{-0.2cm}
    \centering
    \includegraphics[width=0.95\linewidth]{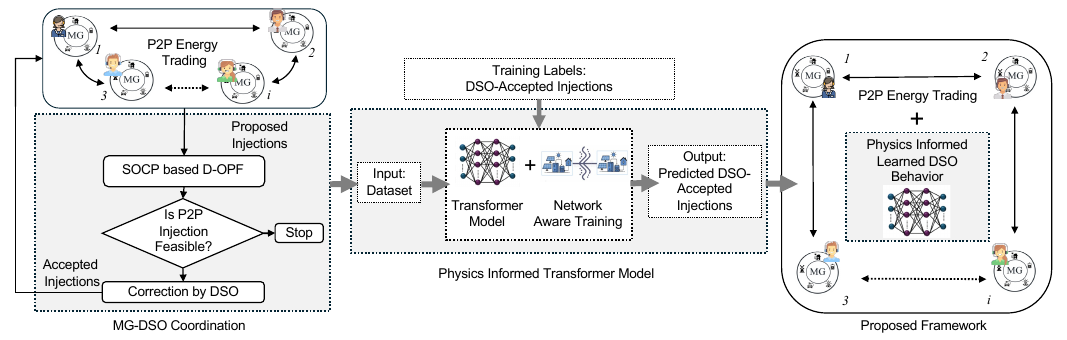}
    \caption{Workflow of the proposed physics informed learning-augmented P2P trading framework}
    \label{fig:workflow}
\end{figure*} 
These limitations motivate the development of a coordination architecture for P2P energy trading that retains distributed autonomy and feasibility while remaining suitable for near real-time operation. To this end, we propose a physics informed learning-augmented P2P framework that enables MGs to internalize the DSO's response within their market decisions. A supervised Transformer-based model is trained to learn the DSO's response while embedding the network constraints into the training objective. The learned network intelligence is subsequently integrated into the P2P market clearing algorithm, enable MGs to locally assess the DSO response without sharing detailed operational information and thereby making the P2P market clearing smarter and reducing the computational burden on the DSO. To the best of our knowledge, learning-based approaches have not been explored for explicitly modeling MG--DSO coordination in network-aware P2P energy trading, although prior works have applied machine learning primarily at the market layer \cite{9596598,9468648}.
\vspace{-0.4cm}  
\subsection{Proposed Learning-augmented P2P Framework}
The workflow of the proposed framework is illustrated in Fig.~\ref{fig:workflow}. The system consists of multiple MGs engaged in P2P energy trading, which are interconnected through a distribution network at distinct nodes. In the baseline bilevel architecture (leftmost block), MGs execute a market-clearing algorithm (usually a game-theoretic approach) to generate net injections, which are iteratively validated and corrected by the DSO using OPF-based feasibility assessment as in \cite{zou2022peer,yan2020distribution}. To capture the DSO behavior, a dataset is generated under diverse operating conditions, including variations in renewable generation, load demand, and network parameters. Based on this dataset, a supervised transformer-based regression model (middle block) is trained to learn the mapping of the proposed injections to DSO-accepted injections. The distribution network operating constraints are incorporated into the training objective through a constraint-violation penalty. In the proposed framework (rightmost block), the learned model (which is provided by the DSO to the market clearing layer) is embedded directly within P2P market clearing algorithm, allowing MGs to proactively align their decisions with network constraints and DSO's economic objectives, enabling an intelligent decision making without computationally heavy iterative DSO intervention.
\vspace{-0.4cm}
\subsection{Contribution}
Building upon the aforementioned framework, the key contributions of this work are summarized as follows:
\begin{enumerate}
    \item A supervised transformer-based regression model with a two-layer encoder architecture is developed to learn the nonlinear mapping between the proposed net injections generated by the MG-level P2P market and the DSO-accepted injections, while embedding distribution-network feasibility directly within the training objective through  constraint-violation penalties.
    \item A physics informed learning-augmented MG–DSO coordination architecture is developed, wherein the P2P market clearing algorithm embeds predictive network awareness, and MGs proactively align their trading decisions, reducing reliance on iterative DSO validation.
    \item The proposed framework has been demonstrated to enhance trade feasibility, while substantially reducing network constraints violations, lowering computational and communication overhead, and reducing DSO burden compared to the bilevel architecture.

\end{enumerate}
The organization of the paper is outlined as follows. Section \ref{section2} presents MG P2P modeling, while section \ref{section3} describes DSO network formulation. Section \ref{section4} details the ML model training process, and embedding the learned model in MG P2P layer. Section \ref{section5} explains the test system, dataset and section \ref{section6} describes the results. Section \ref{section7} concludes the paper.
\section{P2P Energy Trading Modeling}
\label{section2}
This section presents the microgrid operation model, including load, diesel generators (DGs), solar, battery energy storage systems (BESS), and the P2P market-clearing mechanism.
\subsection{Mathematical Modeling of Microgrid}
\label{MGMATHS}
 Let's consider a system of $\mathcal{M}$ MGs, indexed by $i \in \{1, \ldots, \mathcal{M}\}$ over a time horizon of $T$ hours with the time steps $t \in \{1, \ldots, T\}$. Each MG maximizes utility and trading profit while minimizing generation and storage costs. The net payoff function $({\mathcal{W}} _{i,t}) $ for MG $i$ at time $t$ is as follows:
\begin{equation}
     U(\mathcal{L}_{i,t})+\pi_t(\mathcal{P}^S_{i,t}-\mathcal{P}^B_{i,t})-\mathcal{C}(\mathcal{G}_{i,t}) -\lambda_{d}(\mathcal{P}^{dch}_{i,t}+\mathcal{P}^{ch}_{i,t})
     \label{1}
\end{equation}
where, $U(\mathcal{L}_{i,t})$ is the utility function, $\pi_t$ is the P2P cleared price, $\mathcal{P}^{S}_{i,t},\mathcal{P}^{B}_{i,t}$ is the total power sold and bought in P2P market, $\mathcal{C}(\mathcal{G}_{i,t})$ is the cost of the generation and  ($\lambda_{d}(\mathcal{P}^{dch}_{i,t}+\mathcal{P}^{ch}_{i,t}))$ is the battery degradation cost. 
The net power exchange $(\mathcal{P}^{net}_{i,t})$ at time $t$ is given by: 
\begin{equation}
     \mathcal{P}^{net}_{i,{t}} = \mathcal{G}_{i,t} + \mathcal{R}_{i,t} + \mathcal{P}^{dch}_{i,t} - \mathcal{L}_{i,t} - \mathcal{P}^{ch}_{i,t}
     \label{2}
\end{equation}
\textit{If} $ \mathcal{P}^{net}_{i,t}>0$, its a seller, then,
\begin{equation*}
    \mathcal{P}^S_{i,t}=\mathcal{P}^{net}_{i,t}, \mathcal{P}^B_{i,t} =0
\end{equation*}
\textit{If} $ \mathcal{P}^{net}_{i,t}<0$, its a buyer, then, 
\begin{equation*}
    \mathcal{P}^B_{i,t}=-\mathcal{P}^{net}_{i,t}, \mathcal{P}^S_{i,t} =0
\end{equation*}

\subsubsection{Utility function}
A utility function quantifies the satisfaction that the agent receives from consuming energy. A piecewise utility function (\ref{3}) is adopted \cite{samadi2010optimal}, enforcing demand to operate strictly within the prescribed bounds. 
\begin{equation}
U(\mathcal{L}_{i,t}) = 
\left\{
\begin{array}{ll}
\alpha_i \cdot \mathcal{L}_{i,t} - \beta_i \cdot \mathcal{L}_{i,t}^2, & \text{if } \mathcal{L}_{i,t} \leq \dfrac{\alpha_i}{2\beta_i} \\
\dfrac{\alpha_i^2}{4\beta_i}, & \text{if } \mathcal{L}_{i,t} \geq \dfrac{\alpha_i}{2\beta_i}
\end{array}
\right.
\label{3}
\end{equation}
\begin{equation}
    \mathcal{L}_{i,t}^{\text{min}} \leq  \mathcal{L}_{i,t} \leq \mathcal{L}_{i,t}^{\text{max}}
    \label{4}
\end{equation}
A higher $\alpha_i$  (marginal utility coefficient) increases the initial slope of the utility curve, indicating a greater benefit from energy consumption at lower demand levels. In contrast, $\beta_i$  (diminishing return coefficient) governs the saturation behavior of the utility function. Smaller values of $\beta_i$ result in a higher achievable utility, while larger values lead to faster diminishing returns and earlier saturation, as illustrated in Fig.~\ref{fig:utility}.
\subsubsection{Cost of generation}
Generator costs follow a quadratic model (\ref{5}) and its dispatch is constrained by the operating bounds (\ref{6}).
\begin{equation}
    \mathcal{C}(\mathcal{G}_{i,t}) =   a_i \cdot \mathcal{G}_{i,t}^{2} + b_i \cdot \mathcal{G}_{i,t} 
    \label{5}
\end{equation}
\begin{equation}
    0 \leq  \mathcal{G}_{i,t} \leq \mathcal{G}_{i,t}^{\text{max}}
    \label{6}
\end{equation}
where, $a_i$ and $b_i$ denotes the quadratic and linear cost coefficients respectively.
\subsubsection{Battery Energy Storage System (BESS)}
The BESS operation is described through SoC dynamics (\ref{7}) and its associated limits (\ref{8}) (\ref{10}), as formulated below.
\vspace{-0.3cm}
\begin{equation}
\begin{aligned}
SOC_{i,t+1} &= SOC_{i,t} + \eta_c \mathcal{P}_{i,t}^{{ch}}\Delta t
            - \frac{1}{\eta_d} \mathcal{P}_{i,t}^{{dch}}\Delta t\\
\end{aligned}
\label{7}
\end{equation}
\begin{equation}
    0 \leq \mathcal{P}_{i,t}^{{ch}} \leq \mathcal{P}_{i,max}, \quad 0 \leq \mathcal{P}_{i,t}^{dch} \leq \mathcal{P}_{i,max}
    \label{8}
\end{equation}
\begin{equation}
    SOC_{i}^{\min} \leq SOC_{i,t} \leq SOC_{i}^{\max}
    \label{10}
\end{equation}
Battery storage introduces inter-temporal, path-dependent coupling, eliminating the possibility of closed-form solutions unlike \cite{tushar2019grid, ullah2021peer}.
Accordingly, the overall objective of MG is defined as follows.
\vspace{-2mm} 
\begin{equation}
    \max_{\mathcal{L}_{i,t},\, \mathcal{G}_{i,t},\, \mathcal{R}_{i,t},\, \mathcal{P}^{dch}_{i,t},\, \mathcal{P}^{ch}_{i,t}}\hspace{1mm} {\mathcal{W}} _{i,t}
    \label{mg}
\end{equation}
\begin{equation*}
 s.t. \hspace{1mm}  (\ref{2})- (\ref{10})
 \end{equation*}
 \vspace{-10mm}
\begin{figure}[]
    \centering
    \includegraphics[width=\linewidth]{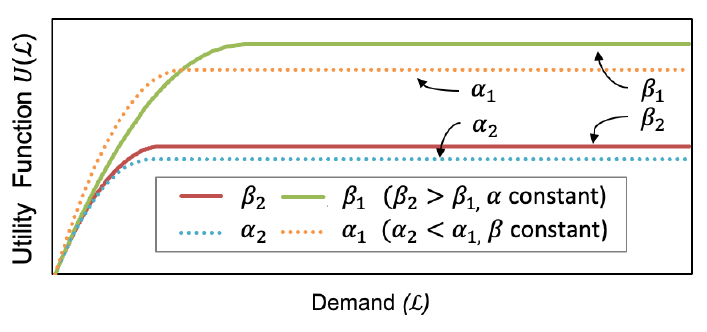}
    \vspace{-6mm}
    \caption{Utility function plot}
    \label{fig:utility}
    \vspace{-3mm}
\end{figure}
\vspace{-2mm}
\subsection{P2P Market Clearing Algorithm}
The P2P market clearing is formulated as a static non-cooperative Nash equilibrium game, where no microgrid can improve its payoff through unilateral deviation \cite{zhu2022nash}.
A strategy profile $P^{*}=P^{*}_{1}, P^{*}_{2}, P^{*}_{3}....,P^{*}_{n}$ is considered as nash equilibrium if, for each player $i$, the following condition hold:
\begin{equation}
    \mathcal{W}_{i}(P_{i}^{*},P_{-i}^{*}) \geq \mathcal{W}_{i}(P_{i},P_{-i}^{*}) \hspace{4mm} \forall P_i \in S_i
    \label{12}
\end{equation}
where, $\mathcal{W}i$ denotes the payoff of player $i$; $P_i^*$ and $P{-i}^*$ represent the equilibrium strategies of player $i$ and the other players, respectively; $P_i$ is any feasible strategy of player $i$, and $S_i$ is the corresponding strategy set. Each microgrid solves its local optimization, as formulated in (\ref{mg}), after which total supply and demand are aggregated to update the market price. The iterations continue until the supply–demand mismatch converges to equilibrium. 

\section{DSO-Network Formulation }
\label{section3}
This section presents the distribution system model and the D-OPF formulation used by the DSO to enforce network constraints and assess feasibility.
\subsection{Distribution Network Modeling}
\begin{figure}[h!]
\vspace{-0.4cm}
    \centering
    \includegraphics[width=0.5\textwidth]{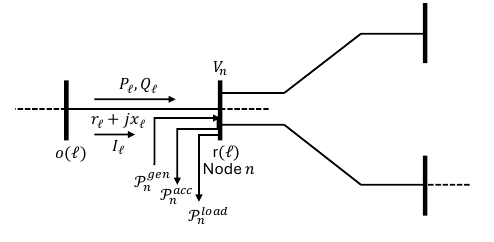}
    \caption{Representation of radial distribution network under consideration}
    \label{fig:radialnetwork}
    \vspace{-0.5cm}
\end{figure}

A radial distribution network, as illustrated in Fig. \ref{fig:radialnetwork}. It can be modeled as a graph $\Gamma(\mathcal{N}, \mathcal{L})$, where $\mathcal{N}$ denotes the set of nodes and $\mathcal{L}$ denotes the set of lines. For any node $n \in \mathcal{N}$, its parent and downstream (child) nodes are identified through the lines $\ell \in \mathcal{L}$ that are incident on it. The parent node corresponding to $n$ is identified as the originating (sending-end) node $o(\ell)$ of the associated line $\ell$. In a similar manner, the set of child nodes associated with node $n$ comprises the receiving-end nodes $r(\ell)$ of the lines $\ell$ originating from $n$ in the downstream direction. Owing to the radial structure, a node can be connected to downstream nodes but is associated with only a single upstream node. The voltage at each node $n \in \mathcal{N}$ is characterized by its magnitude $V_n$, which is constrained within the permissible limits $[\underline{V}_n, \overline{V}_n]$, where $\underline{V}_n$ and $\overline{V}_n$ denote the lower and upper voltage bounds, respectively. We define $v_n = V_n^2$ in order to use linear operators. Note that $v_n$ is also bounded by $\underline{v_n} =(\underline{V_n})^2$ and $\overline{v_n} =(\overline{V_n})^2$. Similarly, current $I_{\ell}$ flows through line $\ell$, where $\underline{I_{\ell}}$ and $\overline{I_{\ell}}$ define the admissible bounds on the current. Here, $c_{\ell} = |I_{\ell}|^2$, where $c_{\ell}$ is bounded by $\underline{c_{\ell}} =|\underline{I_{\ell}}|^2$ and $\overline{c_{\ell}} =|\overline{I_{\ell}}|^2$. Here,
$r_{\ell}$ and $x_{\ell}$ represent the resistance and reactance of line $\ell \in \mathcal{L}$, respectively. 
\subsection{Distribution Optimal Power Flow (D-OPF)}
\label{DOPF}
Network feasibility is enforced by the DSO through a D-OPF formulation using the branch flow model \cite{low2014convex}, while minimizes line losses and penalizing excessive correction. Objective function is defined as,   
\begin{equation}
 \mathcal{F}_{\ell,n} = c^{ls} \sum_{\ell} r_{\ell} c_{\ell}  + \lambda_{corr} \sum_n (\mathcal{P}^{acc}_n - \mathcal{P}^{net}_{n})^2
\label{13}
\end{equation}
where, $c^{ls}$ is the cost coefficient for the line losses, $\lambda_{corr}$ is a penalty factor for correction, and $\mathcal{P}^{net}_{n}, \mathcal{P}^{acc}_n$ are the net and accepted P2P injections at node $n$. The power flow and network constraints are defined as follows.
\begin{equation}
    \sum_{\ell|o(\ell)=n} P_{\ell}- P_{\ell|r(\ell)=n}  = \mathcal{P}_n^{gen}  - \mathcal{P}_n^{load} - \mathcal{P}_n^{acc} - r_{\ell}c_{\ell}
    \label{14}
\end{equation}
\begin{equation}
    \sum_{\ell|o(\ell)=n} Q_{\ell}-Q_{\ell|r(\ell)=n}  = Q_n^{gen} - Q_n^{load} -x_{\ell}c_{\ell}
    \label{15}
\end{equation}
\begin{equation}
    v_{o(\ell)} - v_{r(\ell)} = 2 \left( r_{\ell} P_{\ell} + x_{\ell} Q_{\ell} \right) - \left( r_{\ell}^2 + x_{\ell}^2 \right) c_{\ell}
    \label{16}
\end{equation}
\begin{equation}
    c_{\ell} = \frac{P_{\ell}^2 + Q_{\ell}^2}{v_{n}}
    \label{17}
\end{equation}
\begin{equation}
    \underline{v}_{n} \leq v_{n} \leq \overline{v}_{n}\hspace{3mm} \text{and} \hspace{3mm} 0 \leq c_{\ell} \leq \overline{c}_{\ell}
    \label{18}
\end{equation}
\begin{equation}
      -\,\overline{\mathcal{P}}_n^{acc} \;\le\; \mathcal{P}_n^{acc} \;\le\; \overline{\mathcal{P}}_n^{acc}  
      \label{19}
\end{equation}
\begin{equation}
      0 \;\le\; \mathcal{P}_n^{gen} \;\le\; \overline{\mathcal{P}}_n^{gen} \hspace{3mm} \text{and} \hspace{3mm} 0 \;\le\; Q_n^{gen} \;\le\; \overline{Q}_n^{gen} 
      \label{20}
\end{equation}
The active and reactive power balance at each node $n$ is captured by \eqref{14} and \eqref{15}, respectively.
Equation \eqref{16} describes the voltage drop across a branch, while \eqref{17} defines the branch current magnitude. Equation \eqref{18} imposes the voltage limits at node $n$ and current limits on branch $l$. Finally, \eqref{19} and \eqref{20} represent the P2P injection limit ($\mathcal{\bar{P}}_n^{acc}$) and the generation capacity constraints, respectively. Due to constraint (\ref{17}), the problem is non-convex in nature, to enforce convexity of the problem, the constraint (\ref{17}) is relaxed to (\ref{21}), which makes the problem convex in form of second-order cone programming (SOCP) \cite{farivar2013branch}.
\begin{figure*}[t]
    \centering
    \includegraphics[width=0.92\linewidth]{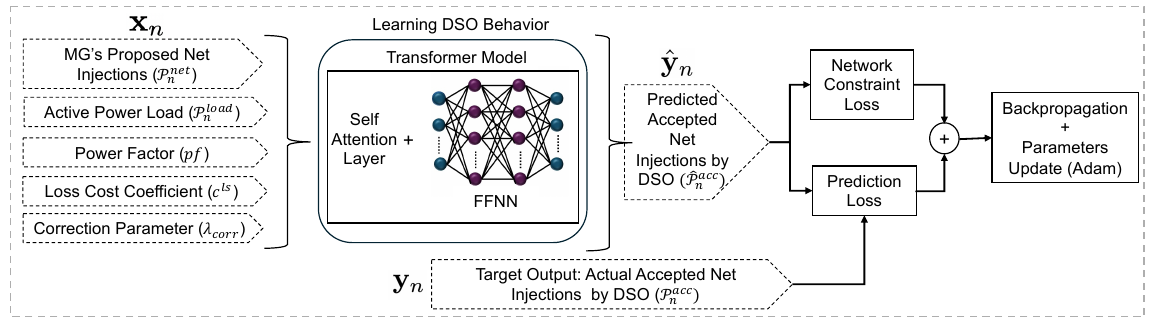}
    \caption{Training process of the physics informed learning - augmented model}
    \label{fig:neural}
\end{figure*}
\begin{equation}
   c_{\ell} \geq \frac{P_{\ell}^{2} + Q_{\ell}^{2}}{v_n} \Rightarrow \left\|\begin{array}{c}
2 P_{\ell} \\
2 Q_{\ell} \\
v_n-c_{\ell}
\end{array}
\right\|_2 \leq v_n+c_{\ell}
   \label{21}
\end{equation}
Convex relaxation enables global-optimality checks that traditional non-linear OPF cannot. If exact, it certifies the global optimum; if not, it provides a lower bound indicating how close any solution is to optimal. An infeasible relaxation also certifies that the original OPF is infeasible \cite{low2014convex}.
Accordingly, the overall objective of DSO is defined as follows.
\begin{equation}
\min_{v_{n},\, c_{\ell},\, P_{\ell},\, Q_{\ell},\, \mathcal{P}^{gen}_{n},Q^{gen}_{n},\mathcal{P}_n^{acc} } \mathcal{F}_{\ell,n}
\label{13}
\end{equation}
\begin{equation*}
 s.t. \hspace{1mm} (\ref{14}-\ref{16}), (\ref{18})-\ref{21})
\end{equation*}

\section{Learning Distribution System Operator Behavior}
\label{section4}
In a typical bilevel architecture (as explained in Fig. \ref{fig:workflow}), the DSO performs a feasibility assessment using D-OPF and suggests corrections to the net P2P injection determined by the market layer. However, in the proposed approach, we intend to predict the DSO response a priori using a physics informed learning-augmented model, whose training process is shown in Fig.~\ref{fig:neural}. 
\subsection{Physics Informed Transformer Model}
\label{subsectiontrans}
\begin{figure}[h!]
\vspace{-0.5cm}
    \centering
    \includegraphics[width=\linewidth]{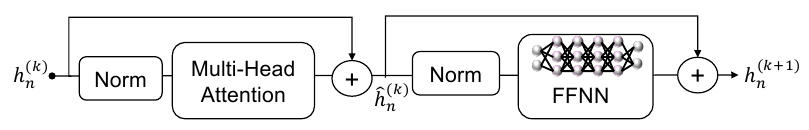}
    \caption{Transformer encoder layer}
    \label{fig:encoderlayer}
\end{figure}

The input vector, $\mathbf{x}_n$, is processed through self-attention layers of a transformer encoder–based model \cite{vaswani2017attention} to capture spatial dependency and interdependency among $n$ buses and $\mathcal{M}$ microgrids, followed by feed-forward networks for nonlinear transformation, to estimate the DSO-accepted injections,$\mathbf{y}_n$, with predictions denoted as $\mathbf{\hat{y}}_n$. The input feature vector for each node $n$ is defined as $\mathbf{x}_n = [\bar{\mathbf{x}}_n,\ \mathbf{g}]$, where, $\quad \bar{\mathbf{x}}_n=  [P_n^{net}, P_n^{load},pf] \quad \text{and} \quad \mathbf{g} = [c^{\text{ls}}, \lambda_{\text{corr}}].$ A 5-dimensional physical feature is projected into a 64-dimensional space through a learnable embedding layer. Mathematically, this operation is defined as:
\begin{equation}
\mathbf{h}_n^{(0)} = \mathbf{W}\,\mathbf{x}_n + \mathbf{b},
\qquad \mathbf{h}_n^{(0)} \in \mathbb{R}^{64}
\end{equation}
The embedded bus representations are processed through a stack of $K=2$ transformer encoder layers. Each encoder employs a self-attention mechanism to capture dependencies. The resulting features are further refined through a feed-forward network, enabling the model to learn complex nonlinear relationships among the system variables. Let $h_n^{(k)}$, denote the input to the $k^{th}$ layer. Each encoder is designed as illustrated in Fig.~\ref{fig:encoderlayer} \cite{devlin2019bert}. 
Each transformer encoder layer processes the input representation $\mathbf{h}_n^{(k)}$ through layer normalization, multi-head attention (MHA), and feed-forward neural network (FFNN) operations with residual connections. Specifically, layer normalization is first applied \cite{ba2016layer}, and is given by $\tilde{\mathbf{h}}_n^{(k)} = \mathrm{LN}\!\left(\mathbf{h}_n^{(k)}\right)$.
\subsubsection{Multi-Head Self Attention}
For self-attention, the query $(\mathbf{Q}_n = \tilde{\mathbf{h}}_n^{(k)} \mathbf{W}_Q)$, key $(\mathbf{K}_m = \tilde{\mathbf{h}}_m^{(k)} \mathbf{W}_K)$, and value $(\mathbf{V}_m = \tilde{\mathbf{h}}_m^{(k)} \mathbf{W}_V)$ vectors are computed as
\cite{dosovitskiy2020image}, where $\mathbf{W}_Q,\mathbf{W}_K,\mathbf{W}_V $ are the learnable projection matrices. \\
The attention weights and the self-attention output for bus $n$ are given by:
\begin{equation}
\alpha_{n,m}
=
\mathrm{softmax}_{m}
\!\left(
\frac{\mathbf{Q}_n \mathbf{K}_m^{\top}}{\sqrt{d_k}}
\right),
\mathbf{SA}_n^{(k)} = \sum_{m} \alpha_{n,m}\,\mathbf{V}_m 
\end{equation}
where, $d_k$ is the key dimension and $m \in \mathcal{N}$ denotes the index of all buses in the network over which attention is computed.  
In the multi-head setting, attention outputs are concatenated and linearly projected to obtain $MHA(\tilde{\mathbf{h}}_n^{(k)})$. 
The output $\hat{\mathbf{h}}_n^{(k)}
=
\mathbf{h}_n^{(k)}
+
\mathrm{MHA}\!\left(\tilde{\mathbf{h}}_n^{(k)}\right)$ is combined with the input via a residual connection, followed by layer normalization ($\tilde{\mathbf{h}}_n^{(k)}
=
\mathrm{LN}\!\left(\hat{\mathbf{h}}_n^{(k)}\right)$).
\subsubsection{Feed Forward Neural Network (FFNN)}
A two-layer FFNN with GELU activation is used for nonlinear feature transformation at each bus.
\begin{equation}
\mathrm{FFNN}\!\left(\tilde{\mathbf{h}}_n^{(k)}\right)
=
\mathbf{W}_2\,
\phi\!\left(
\mathbf{W}_1\,\tilde{\mathbf{h}}_n^{(k)} + \mathbf{b}_1
\right)
+
\mathbf{b}_2 
\end{equation}
where, $\mathbf{W}_1 \in \mathbb{R}^{128 \times 64}, \mathbf{W}_2 \in \mathbb{R}^{64 \times 128}$ and, $\phi(\cdot)$, denotes the GELU activation function. The output of the encoder layer is obtained via a residual connection.
\begin{equation}
\mathbf{h}_n^{(k+1)}
=
\hat{\mathbf{h}}_n^{(k)}
+
\mathrm{FFNN}\!\left(\tilde{\mathbf{h}}_n^{(k)}\right)
\end{equation}
After passing through all encoder layers, the final bus representations $h_n^{(K)}$ are mapped to the output space using a linear projection. 
\begin{equation}
    \hat{\mathbf{y}}_n= \mathbf{W_0}\mathbf{h}_n^{(K)} +b_0, \quad \mathbf{W_0} \in \mathbb{R}^{1 \times 64}
\end{equation}
Here, $\hat{\mathbf{y}}_n \Rightarrow$ $\hat{\mathcal{P}}_n^{acc}$ denotes the predicted accepted net injections by the DSO. Model training is performed by comparing the $\hat{\mathbf{y}}_n$ with the ground truth target $\mathcal{P}_n^{acc}$, that is, the actual net injections accepted by DSO, obtained through the bilevel baseline architecture explained in Fig. \ref{fig:workflow} . 
After getting the $\hat{\mathbf{y}}_n$, the transformer parameters are updated by minimizing the defined loss functions. The complete set of trainable parameters are denoted by $\boldsymbol{\theta} =
\{\boldsymbol{\theta}_{\text{emb}},
\boldsymbol{\theta}_{\text{att}},
\boldsymbol{\theta}_{\text{ffn}},
\boldsymbol{\theta}_{\text{ln}},
\boldsymbol{\theta}_{\text{out}}\}$
where, $\boldsymbol{\theta}_{\text{emb}}=$ $\{\mathbf{W},\mathbf{b}\}$ denotes the parameters of the input embedding layer;
$\boldsymbol{\theta}_{\text{enc}}=\{\mathbf{W}_Q,\mathbf{W}_K,\mathbf{W}_V,\mathbf{W}_1,\mathbf{b}_1,\mathbf{W}_2,\mathbf{b}_2,\boldsymbol{\gamma}^{(k)},\boldsymbol{\beta}^{(k)}\}$ represents the parameters of the Transformer encoder layers, including the multi-head self-attention, feed-forward network, and layer normalization components;
and $\boldsymbol{\theta}_{\text{out}}=\{\mathbf{W}_o,\mathbf{b}_o\}$ denotes the parameters of the output projection layer. These parameters are jointly optimized through backpropagation using the two loss functions that is prediction loss and network constraint loss. Accordingly, the prediction loss is formulated based on the Mean Squared Error (MSE) criterion as
\begin{equation}
\begin{aligned}
\mathcal{L}_{\mathrm{pred}}(\boldsymbol{\theta})= \frac{1}{N} \sum_{n=1}^{N} \left\| \mathbf{y}_n - \hat{\mathbf{y}}_n \right\|_2^2
\end{aligned}
\label{pred}
\end{equation}
To explicitly incorporate the network feasibility directly into the learning process, the Transformer-predicted accepted net injections are mapped to the corresponding nodal injections of the distribution network. The resulting operating point is evaluated against the DSO network constraints defined in Eqs.~(\ref{14})--(\ref{20}), and any resulting violations are penalized through the following physics-informed constraint loss:
\begin{equation}
\mathcal{L}_{\mathrm{cons}}(\boldsymbol{\theta})
=
\frac{1}{N}
\sum_{n=1}^{N}
\Phi\!\left(\hat{\mathbf{y}}_n\right),
\label{eq:physics_loss}
\end{equation}
where $\Phi(\hat{\mathbf{y}}_n)$ denotes the aggregate constraint-violation function obtained by evaluating the network operating point associated with the predicted accepted injections $\hat{\mathbf{y}}_n$. Specifically, the predicted injections are substituted into Eqs.~(\ref{14})--(\ref{20}) to determine the resulting network state, which is then evaluated against the prescribed voltage and line-flow limits. A value of $\Phi(\hat{\mathbf{y}}_n)=0$ indicates that all the considered network constraints are satisfied, whereas $\Phi(\hat{\mathbf{y}}_n)>0$ represents the magnitude of the resulting constraint violations. Accordingly, the proposed Transformer is trained by minimizing the composite objective 
\begin{equation}
\mathcal{L}(\boldsymbol{\theta})
=
\mathcal{L}_{\mathrm{pred}}(\boldsymbol{\theta})
+
\mathcal{L}_{\mathrm{cons}}(\boldsymbol{\theta}),
\label{eq:total_loss}
\end{equation}
which gives the following optimization problem:
\begin{equation}
\boldsymbol{\theta}^{\star}
=
\arg\min_{\boldsymbol{\theta}}
\left[
\mathcal{L}_{\mathrm{pred}}(\boldsymbol{\theta})
+
\mathcal{L}_{\mathrm{cons}}(\boldsymbol{\theta})
\right].
\label{eq:training_objective}
\end{equation}
The gradients of the composite loss are backpropagated through the Transformer network, thereby updating the model parameters towards both prediction accuracy and distribution-network feasibility \cite{9743327}. The model parameters are updated iteratively using gradient-based optimization with the Adam optimizer \cite{arya2022effective}.

\begin{algorithm}[b]
\caption{Microgrid Optimization Using Learned DSO Response }
\label{alg:MLnash}
\begin{algorithmic}[1]
\STATE Initialization: Set system parameters, $\xi$, $\varepsilon$, $T$, $K_{\max}$.
\FOR{$t = 1$ to $T$}
        \WHILE{$|{\mathcal{P}^{B,acc}_k} - {\mathcal{P}^{S,acc}_k}|\ge  \varepsilon$ \AND $k < K_{\max}$}
            \STATE Step 1: Microgrid optimization
               \FOR{each MG $i = 1,\ldots,\mathcal{M}$}
                  \STATE Solve the optimization of MG $i$ (Subsection \ref{MGMATHS}) incorporating the learned DSO response\\
                  \STATE $\mathcal{W}_{i,t} - \mu \left(\hat{\mathcal{P}}_{i,t}^{acc} - \mathcal{P}_{i,t}^{net}\right)^2$
               \ENDFOR
            \STATE Step 2: Market Clearing Layer
            \STATE Compute total power supply ${\mathcal{P}^{S,acc}_k}$  and total power demand ${\mathcal{P}^{B,acc}_k}$.
            \STATE Evaluate the learned model ($f_{ML}(\mathbf{x}_n)$)  and communicate 
            $\hat{P}_i^{acc,(k+1)}$ to each MG $i$.
            \STATE Step 3: Price update\\
             $\pi_{k+1} = \pi_{k} + \xi\!\left({\mathcal{P}^{B,acc}_k} - {\mathcal{P}^{S,acc}_k}\right)$
        \ENDWHILE 
\ENDFOR
\STATE Output: Equilibrium price and trades for all MG's.
\end{algorithmic}
\end{algorithm}
\vspace{-0.3cm}
\subsection{Incorporating ML model in P2P trading}
Let $\mathcal{F}_{DSO}$ denote the feasible operating region, defined by the SOCP model, presented in Subsection \ref{DOPF}, i.e. the set of all accepted net injections vectors that satisfies the distribution network constraints. For any proposed net power injection by the P2P market layer, $\mathcal{P}_n^{net}$, the DSO-accepted power is obtained as the projection onto the feasible region:
\begin{equation}
    \mathcal{P}_n^{acc} = \Pi_{\mathcal{F}_{DSO}}(\mathcal{P}_n^{net})
\end{equation}  
As explained in Fig. \ref{fig:workflow}, to avoid explicitly solving this projection at each iteration in distributed P2P market structures, a learned mapping is introduced, 
$(f_{ML}(\mathbf{x}_n) \rightarrow \hat{\mathcal{P}}_n^{acc})$, 
which approximates the DSO response based on the input $\mathbf{x}_n$. This learned model is embedded within the market clearing layer (Algorithm~\ref{alg:MLnash}). To evaluate the learned model, the market clearing layer receives the network parameters in real time from the DSO and the net P2P trades for the previous iteration from MGs, and communicate the respective $\hat{\mathcal{P}}_{i,t}^{acc}$ and price values to MGs. The MG embed those values inside the objective function (originally defined in Section \ref{MGMATHS}) as follows:
\begin{equation}
  \widetilde{W}_{i,t}=  \mathcal{W}_{i,t} - \mu \left(\hat{\mathcal{P}}_{i,t}^{acc} - \mathcal{P}_{i,t}^{net}\right)^2
    \label{32}
\end{equation}

where $\mu > 0$ acts as a regularization coefficient that penalizes the deviation between the proposed net power and the ML-predicted network-feasible value. By incorporating this term, the MG internalizes the anticipated network feasibility constraints and substantially reduces the likelihood of generating infeasible operating points. Consequently, the proposed net power $\mathcal{P}_n^{net}$ is driven closer to the feasible region, reducing projection-induced curtailment and improving the efficiency of power exchange. Algorithm~\ref{alg:MLnash} summarizes the overall market-clearing procedure. At each iteration, every MG solves its local optimization problem while accounting for both its economic objective and the ML-based feasibility regularization term. The resulting net-power decisions are iteratively updated through the Nash-equilibrium-based market-clearing process until the prescribed convergence criterion is satisfied. Since the MG strategy sets are nonempty, compact, and convex, and the added regularization term preserves the concavity of the original payoff, so $\widetilde{W}_{i,t}$ stays concave and hence a pure-strategy Nash equilibrium exists in each stage game \cite{rosen1965existence}.

\section{Test System and Dataset}
\label{section5}
The P2P optimization problems of MG, were solved using the Sequential Least Squares Programming (SLSQP)~\cite{nocedal2006numerical}, while the DSO's D-OPF was implemented in CVXPY and solved using Embedded Conic Solver (ECS). The transformer model was trained in scikit-learn using the Adam optimizer. Codes were executed in the Anaconda environment using Spyder IDE with Python version as 3.12.2 on a MacBook Pro (Apple M4, 16 GB RAM, 10 Core GPU) running macOS Sequoia 15.4.1.
\begin{figure}[b]
    \centering
    \includegraphics[width=\linewidth]{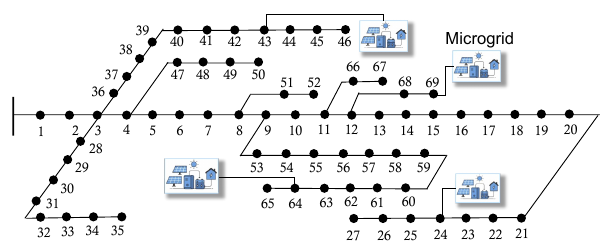}
    \caption{Network topology of IEEE-69 bus system}
    \label{fig:ieee69}
    \vspace{-0.2cm}
\end{figure}
\subsection{Test System}
For the analysis, the modified IEEE 69-bus distribution network \cite{5n30-5402-24} is used as the test system for evaluating the proposed framework as shown in Fig.~\ref{fig:ieee69}. The four microgrids are randomly positioned at Buses 24, 43, 64, and 69 in the network. Each microgrid in the system contains the load, DGs, solar and BESS whose modeling has been done in subsection \ref{MGMATHS}. The MG system parameters is shown in Table \ref{tab:system_parameters}.

For load, the minimum and the maximum demand for each MG is $15$kW and $100$kW respectively. The marginal utility coefficient ($\alpha_i$) has been varying throughout the day between $(0.5-1))$Rs/kWh . The diminishing return coefficient ($\beta_i$) has been varying from $(0.0045-0.003))$ Rs/kWh$^2$ throughout the day. The relationship between $\alpha_i$ and $\beta_i$ and their impact on the utility function are illustrated in Fig.~\ref{fig:utility}. Generator cost coefficients vary across units (cheap to expensive), with linear coefficients in the range 0.079-0.5 Rs/kW and quadratic coefficients 0.0001–0.009 Rs/kW$^2$ \cite{wood2013power}. The learning rate $(\xi)$ for the market clearing is used as 0.005 (to provides the best trade-off between speed and stable convergence). The convergence threshold $(\varepsilon)$ is taken as 0.001.

\begin{table}[t]
\centering
\caption{System parameters of microgrids}
\setlength{\tabcolsep}{10pt}
\renewcommand{\arraystretch}{1.2}
\begin{tabular}{
    >{\arraybackslash}m{4.5cm}|
    >{\centering\arraybackslash}m{2.7cm}
}
\hline
\hline
\multicolumn{2}{c}{Photovoltaic (PV) System} \\
\hline
Nominal PV capacity & 10 kW \\
\hline
\multicolumn{2}{c}{Battery Energy Storage System (BESS)} \\
\hline
Rated power  & 10 kW \\
Minimum/Maximum state of charge   & 4 kWh/20 kWh \\
Charging / discharging efficiency  & 0.95 \\
\hline
\multicolumn{2}{c}{Diesel Generator (DG)} \\
\hline
Maximum generation (MG1, MG3)   & 52 kW \\
Maximum generation (MG2, MG4) & 44 kW \\
\hline
\hline
\end{tabular}
\label{tab:system_parameters}
\end{table}
\subsection{Dataset Generation}
\begin{figure}[h!]
    \centering
    \includegraphics[width=0.97\linewidth]{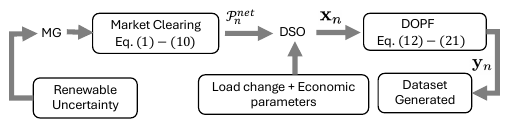}
    \caption{Dataset generation}
    \label{fig:dataset}
\end{figure}

For the machine-learning model, the dataset is generated using the procedure in Fig.~\ref{fig:dataset}. 
\begin{table}[b]
\centering
    \caption{Pearson correlation matrix for input parameters to the transformer model}
    \setlength{\tabcolsep}{6pt}
    \renewcommand{\arraystretch}{1.2}  
    \begin{tabular}{l|ccccc}
    \hline
    \hline
                & $\mathcal{P}_n^{load}$ & $pf$ & $\mathcal{P}^{net}_{n}$ & $c^{ls}$   & $\lambda_{corr}$ \\
    \hline
$\mathcal{P}_n^{load}$ &1& -0.043258&0.000269&0.002366&-0.005227 \\
$pf$ & -0.043258 & 1  & -0.09222 & 0.047382  & -0.04312 \\
$\mathcal{P}^{net}_{n}$& 0.000269 &-0.09222 & 1& -0.006556 & -0.008725 \\
$c^{ls}$& 0.002366 & 0.047382  & -0.006556 & 1 & -0.004120 \\
$\lambda_{corr}$&-0.005227 & -0.04312 & -0.008725 & -0.004120 & 1 \\
    \hline
    \hline
    \end{tabular}
    \label{tab:PCM}
\end{table}
A total of 50,000 scenarios are produced, with an 80–20 train–test split. The input vector, $\mathbf{x}_n$, include the load  $\mathcal{P}_n^{load}$ (obtained through load scaling within $[0.6,1.4]$), power factor $pf\in[0.85,1.0]$ , the net proposed injections by P2P market $\mathcal{P}^{net}_n$ (obtained through the market clearing process under solar uncertainty of $[-20\%,20\%]$), the loss-cost coefficient $c^{ls}\in[3,10]~\mathrm{Rs/kW^2}$, and the correction parameter $\lambda_{\mathrm{corr}}\in[3,10]~\mathrm{Rs/kW^2}$ (set to match the typical price band of the Indian Energy Exchange). Finally, the output $\mathbf{y}_n$ ($\mathcal{P}^{acc}_n$) is obtained by solving the DOPF. The pearson correlation matrix (Table~\ref{tab:PCM}) shows low inter-variable correlation. 
 
\section{Results and Discussion}
\label{section6}
This section evaluates the learning model’s performance and validates the performance of the proposed approach through a set of case studies, explained in the late subsection.
\subsection{Learning Model Prediction}
This subsection evaluates the physics informed learning augmented model for predicting the DSO-accepted net power injections ($\mathcal{P}_n^{acc}$), enabling network-aware MG optimization. Subsection \ref{subsectiontrans} gives the detailed modeling of the learning model. 
\begin{figure}[h!]
    \vspace{-0.2cm}
    \centering
    \includegraphics[width=0.94\linewidth]{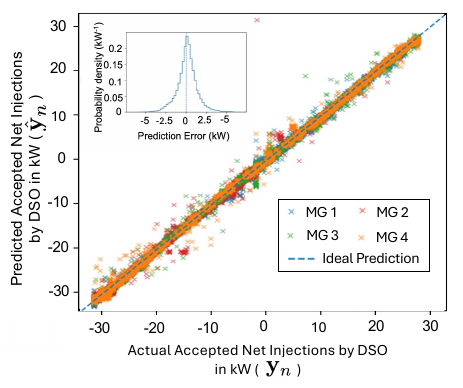}
    \caption{Parity plot}
    \label{fig:parity}
    \vspace{-0.2cm}
\end{figure}

The parity plot  as shown in Fig.~\ref{fig:parity}, shows that most predicted DSO-accepted net power injections lie in close proximity to the 45° reference line, indicating strong agreement with ground-truth D-OPF decisions for both injection and withdrawal regimes. 
The compact clustering around the diagonal reflects low prediction bias and high regression accuracy, while the inset error distribution is sharply centered around zero, confirming small and approximately unbiased errors. 

\begin{figure}[h!]
    \centering
    \includegraphics[width=0.88\linewidth]{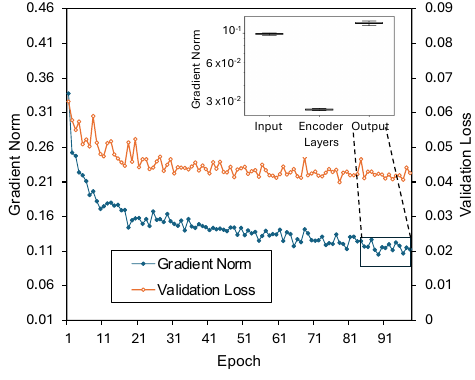}
    \caption{Gradient norm and validation loss}
    \label{fig:grad_loss}
    \vspace{-0.3cm}
\end{figure}

The training dynamics in Fig.~\ref{fig:grad_loss} further demonstrates stable learning behavior, as the gradient norm decreases and stabilizes over epochs, indicating the absence of exploding or vanishing gradients. Similarly, the validation loss exhibits a rapid initial decline followed by smooth convergence, suggesting good generalization. The inset box plot highlights the  balanced gradient flow across the embedding, encoder, and output layers, confirming stable convergence of the model. 

To demonstrate the effectiveness of incorporating network-aware information during training (physics informed), three learning strategies are compared: (a) where the physics-informed network-constrained loss is incorporated into the training objective (Eq.\ref{eq:physics_loss}) (b) where the transformer is trained using only the prediction loss (Eq.\ref{pred}), and (c) feedforward networks. Table \ref{tab:ml_performance} shows how incorporating the network-constrained loss further reduces both $MAE$ and $RMSE$ while increasing the coefficient of determination to $R^2=98.51\%$  has the ability to learn the underlying distribution network constraints (Subsection \ref{DOPF}).
\begin{table}[h!]
\centering
\caption{Performance Comparison of the Learning Model}
\label{tab:ml_performance}
\setlength{\tabcolsep}{5pt}
\renewcommand{\arraystretch}{1.2}

\begin{tabular}{
    p{3.8cm}|
    >{\centering\arraybackslash}m{1.5cm}
    >{\centering\arraybackslash}m{1.6cm}
    >{\centering\arraybackslash}m{0.7cm}
}
\hline\hline
\multicolumn{1}{c|}{Model} &
MAE (MW) &
RMSE (MW) &
$R^2$ \\ 
\hline
a) Physics Informed Transformer
& $1.53\times10^{-4}$
& $3.24\times10^{-4}$
& .9851 \\
b) Transformer 
& $3.23\times10^{-4}$
& $8.19\times10^{-4}$
& .9484 \\
c) Feedforward Neural Network
& $4.09\times10^{-4}$
& $9.51\times10^{-4}$
& .9216 \\

\hline\hline
\end{tabular}
\vspace{-0.2cm}
\end{table}
\vspace{-0.3cm}
\subsection{Case Studies}
\begin{figure}[h!]
    \centering
    \includegraphics[width=\linewidth]{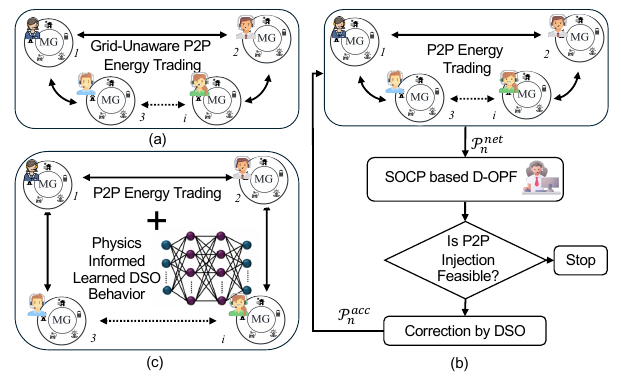}
    \caption{Illustration of the three case studies, where each case is a P2P energy trading mechanism with: (a) \textit{Case 1}: grid-unawareness, (b) \textit{Case 2}: bilevel DSO interaction, and (c) \textit{Case 3}: physics informed learning augmented DSO-awareness.}
    \label{fig:case_studies}
    \vspace{-0.4cm}
\end{figure}

To evaluate the effectiveness of the proposed framework at the system level, three case studies have been designed. The key features of each case study are summarized as follows.
\begin{enumerate}
    \item \textit{Case 1 (Grid-unaware):} Distributed P2P energy trading among MGs, driven solely by local economic objectives without considering network constraints, as illustrated in Fig.~\ref{fig:case_studies}(a).

    \item \textit{Case 2 (DSO-bilevel):} Conventional bilevel DSO coordination architecture, where P2P outcomes are iteratively validated and adjusted via D-OPF to enforce network feasibility (Fig.~\ref{fig:case_studies}(b)), used in (e.g., \cite{zou2022peer,yan2020distribution}). This case serves as the state-of-the-art baseline for assessing the proposed framework's performance.

    \item \textit{Case 3 (Proposed):} Proposed learning-based coordination framework, where the DSO’s operational behavior is embedded into P2P market clearing, allowing MGs to internalize network feasibility, as shown in Fig.~\ref{fig:case_studies}(c).
\end{enumerate}

The performance of the three P2P case studies is evaluated over a 24-hour horizon based on four aspects: (i) power traded in P2P energy trading,(ii) MG's economic performance, (iii) network constraint violations,  and (iv) communication and computational overhead.
\subsubsection{Power Traded in P2P Energy Trading}
\begin{figure}[b]
\vspace{-0.4cm}
    \centering
    \includegraphics[width=0.9\linewidth]{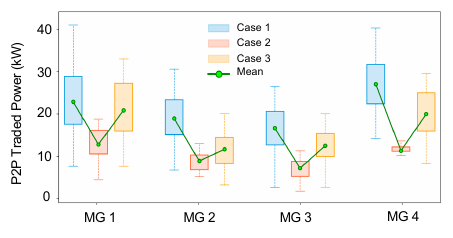}
    \caption{Amount of power traded in P2P market in all Cases}
    \label{fig:p2ppower}
\end{figure}
There is a significant influence of network constraints on the magnitude of the power traded in the P2P market, as shown in Fig.~\ref{fig:p2ppower}. The box plot indicates a noticeable reduction in the magnitude of P2P power exchanges among MGs, when OPF-based DSO coordination \textit{(Case 2)} is enforced, highlighting the restrictive effect of distribution network limits on excessive trading. Amount of P2P power traded is reduced by 34.32\%, 47.55\%, 51.83\%, and 49.56\% for MG1–MG4, respectively, compared to \textit{Case 1}. In contrast, the proposed learning-based framework \textit{(Case 3)} significantly increases the amount power traded in the P2P market, with improvements of 37.15\%, 18.52\%, 56.68\%, and 48.51\% for MG1–MG4, respectively, compared to \textit{Case 2}. This improvement demonstrates that the proposed framework effectively restores trading flexibility while maintaining network feasibility, unlike \textit{Case 2}, which relies on reactive, post-correction feasibility enforcement. By embedding the predicted DSO response into P2P market clearing, the framework enables proactive feasibility-aware decisions, reducing rejection losses and increasing successful transactions.  
\begin{figure}[t]
    \vspace{-0.3cm}
    \centering
    \includegraphics[width=\linewidth]{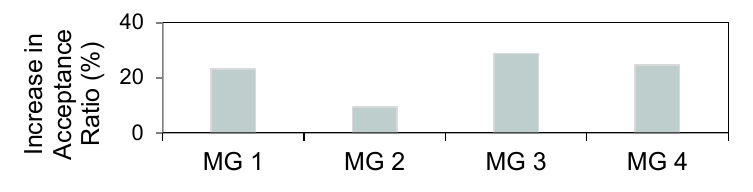}
    \caption{Increase in the acceptance ratio of \textit{Case 3} compared with \textit{Case 2}}
    \label{fig:acceptance}
\end{figure}
\begin{table*}[t]
\centering
\caption{Economic performance comparison of MG1 in all 3 Cases in 24-hour horizon}
\label{tab:mg_profits}
\renewcommand{\arraystretch}{1.3}
\setlength{\tabcolsep}{10pt}
\begin{tabular}{
p{3.1cm}|
>{\centering\arraybackslash}p{2.1cm}|
>{\centering\arraybackslash}p{2.6cm}
>{\centering\arraybackslash}p{2.2cm}
>{\centering\arraybackslash}p{2.0cm}
>{\centering\arraybackslash}p{1.8cm}
}
\hline
\hline
\multicolumn{1}{c|}{Case} & Total MG Payoff $\mathcal{W}_{i,t}$ (Rs) 
 & Net Trading Revenue $\pi_t(P_{{i,t}}^S - P_{{i,t}}^B)$ (Rs)
 & P2P Power Traded $P_{i,t}^{net}$ (kW)
 & Generation Cost $C(\mathcal{G}_{i,t})$ (Rs)
 & Utility Term $U(\mathcal{L}_{i,t})$ (Rs)\\
\hline
Case 1 (Grid-Unaware) & 1981.86 & 2677.877 & 551.564 & 904.0016 & 207.993\\

Case 2 (DSO-Bilevel)& 1106.02 &1876.060  & 357.579 & 950.041 & 180.001 \\

Case 3 (Physics Informed) & 1537.43 & 2267.570 & 472.432 & 925.160   & 195.023 \\
\hline
\hline
\end{tabular}
\end{table*}

\subsubsection{MG Economic Performance}
As shown in Table \ref{tab:mg_profits}, the payoff of MG1 declines by approximately 44.19\% in \textit{Case 2}, compared to \textit{Case 1} (from 1981.86 to 1106.02), due to the impact of network-aware coordination. Similar reductions are observed for the remaining microgrids, with decreases of 37.41\%, 28.24\%, and 27.65\% for MG2–MG4, respectively. Whereas, in \textit{Case 3}, the total payoff of MG1 increases from 1106.02 (\textit{Case 2}) to 1537.43 (\textit{Case 3}), demonstrating a significant recovery in economic performance. 
Similar trends are observed for the remaining microgrids. Table \ref{tab:mg_profit_increase} quantifies the percentage improvement in profit achieved under the proposed framework relative to \textit{Case 2}.

The improvement in economic performance is directly associated with the higher trade acceptance ratio achieved under the proposed learning-augmented framework (\textit{Case 3}) compared to \textit{Case 2}, the percentage increase in the ratio, is shown in Fig.~\ref{fig:acceptance}. The acceptance ratio is defined as the ratio of power traded in a case with respect to \textit{Case 1}. A lower value indicates higher DSO rejection of P2P trades due to network constraints.

\begin{table}[t]
\centering
\caption{Economic recovery in the proposed approach (\textit{Case 3}) for all MGs compared with \textit{Case 2}}
\label{tab:mg_profit_increase}
\footnotesize
\renewcommand{\arraystretch}{1.3}
\setlength{\tabcolsep}{6pt}
\begin{tabular}{lcccc}
\hline
\hline
 & MG1 & MG2 & MG3 & MG4 \\
\midrule
\% Increase in Profit & 39.01\% & 22.32\% & 24.23\% & 26.74\% \\

\hline
\hline
\vspace{-0.7cm}
\end{tabular}
\end{table}
\begin{table}[b]
\vspace{-0.3cm}
\centering
\caption{Execution time}
\label{tab:time}
\renewcommand{\arraystretch}{1.1} 
\begin{tabular}{
>{\centering\arraybackslash}m{2.3cm}
>{\centering\arraybackslash}m{1.5cm}
>{\centering\arraybackslash}m{1.5cm}
>{\centering\arraybackslash}m{1.5cm}
}
\hline
\hline
 & Case 1 & Case 2 & Case 3  \\
\midrule
Time taken (secs) & 52.12 & 349.56 & 67.12 \\
\hline
\hline
\end{tabular}
\end{table}
\begin{table*}[!b]
\centering
\caption{Comparative Analysis of case studies in a time horizon of one year (8760 hours)}
\label{tab:conclusion}
\footnotesize
\renewcommand{\arraystretch}{1.3}
\setlength{\tabcolsep}{8pt}
\centering
\begin{tabular}{m{2.7cm}|cccc}

\hline
\hline
\multicolumn{1}{c|}{Case }
& Total P2P Power Traded (MW) 
& Total MG Payoff (Million Rs)
& Network-Violated (Hrs)
& Time (Minutes) \\
\hline

Case 1 (Grid-Unaware)
& 700.89 (High) 
& 1.605 (High)
& 1942/8760  
& 196.81 (Minimal)\\
   
Case 2 (DSO-Bilevel)
& 368.38 (Reduced by 47.4\%) 
& 1.027 (Degraded by 35.98\%)  
& 0/8760  
& 1105.79 (Iterative) \\
 
Case 3 (Table \ref{tab:ml_performance}-(b)) 
& 538.56 (Improved by 46.20\%) 
&  1.301 (Recovered by 26.58\%) 
&  574/8760 
&  253.21 (Minimal)\\

Case 3 (Table \ref{tab:ml_performance}-(a)) 
& 519.16 (Improved by 40.90\%) 
& 1.254 (Recovered by 22.00\%) 
& 0/8760  
& 272.76 (Minimal) \\
\hline
\hline

\end{tabular}
\end{table*}

\subsubsection{Network Constraint Violations}
\begin{figure}[t]
\vspace{-0.4cm}
    \centering
    \includegraphics[width=0.93\linewidth]{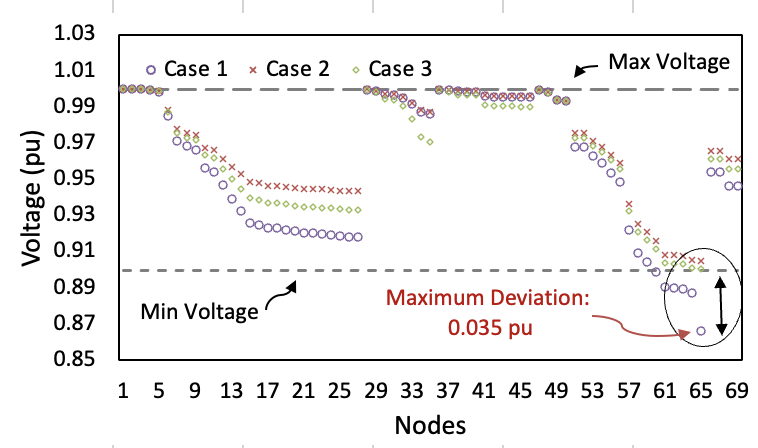}
    \caption{Voltage profile of IEEE-69 bus system (for the time period showing maximum voltage deviation) }
    \label{fig:networkviol}
    \vspace{-0.5cm}
\end{figure}

Among the 24 hourly operating conditions, voltage limit violations are observed during 7/24 hours. Fig.~\ref{fig:networkviol} compares the nodal voltage profiles for the hour exhibiting the maximum voltage deviation under all case studies against the permissible voltage limits of 0.90--1.00 p.u. In \textit{Case 1}, voltage violations are observed, with a maximum deviation of 0.035 pu, indicating infeasible P2P trades. In contrast, in \textit{Case 2}, D-OPF based coordination at the DSO level ensures compliance with voltage constraints across the distribution network, emphasizing the necessity of enforcing network feasibility within decentralized P2P energy markets.
Under the proposed learning-based framework (\textit{Case 3}), voltage profiles remain within permissible limits, and demonstrates that the learned DSO response effectively preserves network feasibility.
 
\subsubsection{Communication and computational overhead}

The average computational burden is quantified in terms of the total execution time required to achieve market convergence over the simulation horizon. Table \ref{tab:time} presents the execution time of one-time P2P trading process in each case. In \textit{Case 1}, the execution time is relatively low, as each microgrid solves only its local optimization problem. In contrast, \textit{Case 2} incurs higher computational and communication overhead due to iterative MG–DSO coordination for feasibility validation. In contrast, the proposed \textit{Case 3} embeds the learned DSO behavior (trained under diverse operating conditions) within the P2P market clearing, enabling network-feasible market outcomes while significantly reducing computational time, communication overhead, thereby minimizes the DSO burden. 

Finally, Table~\ref{tab:conclusion} summarizes the quantitative comparison of the three case studies over a one-year horizon (8760 h). Although \textit{Case 1} achieves the highest traded power and economic transactions, it results in infeasible operation. In contrast, \textit{Case 2} ensures feasibility, but reduces the traded power and economic transactions by 47.4\% and 35.98\%, respectively, while increasing the communication overhead to 1105.79 minutes. The proposed learning-based framework (\textit{Case 3}) is evaluated using two training strategies, as summarized in Table~\ref{tab:ml_performance}: (b) training without the network-constrained loss and (a) training with the physics-informed network-constrained loss. The results show that the physics-informed loss eliminates network violations while balancing market efficiency and secure network operation.

The proposed physics-informed model assumes a fixed network topology during training. Consequently, unforeseen contingencies, including line outages, generator outages, or topology reconfigurations, may require the DSO's intervention. The DSO retains ultimate authority over network operation, exception-triggered verification, continuing to serve as a safety backstop for out-of-distribution conditions.

\section{Conclusion}
\label{section7}
This paper proposes a physics informed learning-augmented, P2P–DSO interface that embeds the DSO behavior within the P2P market clearing process of MGs. It enables MGs to anticipate the DSO's network-feasible decisions while significantly reducing the DSO burden. Quantitative results on the modified IEEE 69-bus system demonstrate that enforcing network feasibility reduces P2P transaction volumes by 47.4\% and decreases the total economic transactions by 35.98\% due to post-market curtailments. In contrast, the proposed framework improves market utilization by 40.90\% and recovers the total economic transactions by 22.00\% while maintaining network feasibility. Furthermore, reduces the centralized coordination and communication overhead, thereby enhancing suitability for near real-time deployment in DER-rich networks. Overall, the proposed framework enables feasible and efficient decentralized P2P market coordination.

\begin{small}
\bibliographystyle{IEEEtran}
\bibliography{references}
\vfill
\end{small}
\end{document}